\documentclass[preprint]{ptephy_v1}

\preprintnumber{KYUSHU-HET-248, OU-HET-1155} 

\usepackage{amsmath} 
\usepackage{amsthm} 
\usepackage{url} 

\usepackage{stmaryrd}

\numberwithin{equation}{section}
\allowdisplaybreaks

\begin{document}

\title{Fractional topological charge in lattice Abelian gauge theory}


\author{Motokazu Abe}
\affil{Department of Physics, Kyushu University, 744 Motooka, Nishi-ku,
Fukuoka 819-0395, Japan}

\author{Okuto Morikawa}
\affil{Department of Physics, Osaka University, Toyonaka, Osaka 560-0043,
Japan}

\author[1]{Hiroshi Suzuki}





\begin{abstract}%
We construct a non-trivial $U(1)/\mathbb{Z}_q$ principal bundle on~$T^4$
from the compact $U(1)$ lattice gauge field by generalizing L\"uscher's
constriction so that the cocycle condition contains $\mathbb{Z}_q$ elements
(the 't~Hooft flux). The construction requires an admissibility condition on
lattice gauge field configurations. From the transition function so
constructed, we have the fractional topological charge that is $\mathbb{Z}_q$
one-form gauge invariant and odd under the lattice time reversal
transformation. Assuming a rescaling of the vacuum angle $\theta\to q\theta$
suggested from the Witten effect, our construction provides a lattice
implementation of the mixed 't~Hooft anomaly between the $\mathbb{Z}_q$
one-form symmetry and the time reversal symmetry in the $U(1)$ gauge theory
with matter fields of charge~$q\in2\mathbb{Z}$ when $\theta=\pi$, which was
studied by Honda and Tanizaki [J. High Energy Phys. \textbf{12}, 154 (2020)] in
the continuum framework.
\end{abstract}

\subjectindex{B01, B02, B06, B31}

\maketitle

\section{Introduction}
\label{sec:1}
As shown in a seminal paper~\cite{Gaiotto:2017yup}, the generalized
symmetries~\cite{Gaiotto:2014kfa} can tell us quite non-trivial information on
the low-energy dynamics of 4D gauge theories through the idea of anomaly
matching~\cite{tHooft:1979rat}; see
Refs.~\cite{McGreevy:2022oyu,Cordova:2022ruw} and references cited therein. In
the present paper, aiming at transparent understanding of the above study in a
fully regularized framework, we construct a non-trivial $U(1)/\mathbb{Z}_q$
principal bundle on~$T^4$ in the compact $U(1)$ lattice gauge
theory.\footnote{Throughout the present paper, we assume that $q$ is a positive
integer.} In the study of the $SU(N)$ gauge theory
in~Ref.~\cite{Gaiotto:2017yup}, the fractionality of the topological charge in
the $SU(N)/\mathbb{Z}_N$ theory and the $\mathbb{Z}_N$ one-form symmetry are
crucially important. The $\mathbb{Z}_N$ one-form gauge transformation can be
interpreted as the action on the transition function in the
$SU(N)/\mathbb{Z}_N$ principal bundle~\cite{Kapustin:2014gua}. The cocycle
condition of the $SU(N)/\mathbb{Z}_N$ principal bundle can contain
$\mathbb{Z}_N$ elements in contrast to the $SU(N)$ bundle, and those
$\mathbb{Z}_N$ elements (the 't~Hooft flux~\cite{tHooft:1979rtg}) are
interpreted as (the gauge-invariant content of) the $\mathbb{Z}_N$ two-form
gauge field transforming under the $\mathbb{Z}_N$ one-form gauge
transformation. These materials are nicely summarized in~Appendix~A
of~Ref.~\cite{Tanizaki:2022ngt}. From this interpretation of the $\mathbb{Z}_N$
one-form gauge transformation, for our motivation, it is natural to consider
the $SU(N)/\mathbb{Z}_N$ principal bundle in lattice gauge theory.

It is well-known that the transition function of the $SU(N)$ principal bundle
on~$T^4$ can be constructed from the $SU(N)$ lattice gauge field by L\"uscher's
method~\cite{Luscher:1981zq}; see
also~Refs.~\cite{Phillips:1986qd,Phillips:1990kj}. In the present paper, we
consider a simpler $U(1)$ gauge theory and generalize L\"uscher's method so
that the cocycle condition contains $\mathbb{Z}_q$ elements; the $\mathbb{Z}_q$
elements are introduced in the transition function and as a loop
in~$U(1)/\mathbb{Z}_q$ and as the loop in $SU(N)/\mathbb{Z}_N$ considered
in~Refs.~\cite{tHooft:1979rtg,vanBaal:1982ag}. Our transition function with the
't~Hooft flux gives rise to the topological charge in the compact $U(1)$
lattice gauge theory that generally takes fractional values. Also, our
topological charge is odd under the lattice time reversal transformation. These
properties are quite analogous to the properties of the $SU(N)/\mathbb{Z}_N$
topological charge that are crucial in the study
of~Ref.~\cite{Gaiotto:2017yup}. In fact, if we assume a rescaling of the
vacuum angle~$\theta$ in the $U(1)$ gauge theory $\theta\to q\theta$ suggested
from the Witten effect~\cite{Witten:1979ey}~(see
Refs.~\cite{Honda:2020txe,Hidaka:2019jtv}), our construction provides a lattice
implementation of the mixed 't~Hooft anomaly between the $\mathbb{Z}_q$
one-form symmetry and the time reversal symmetry in the $U(1)$ gauge theory
with matter fields of charge~$q\in2\mathbb{Z}$ when $\theta=\pi$. This anomaly,
which was studied by Honda and Tanizaki in~Ref.~\cite{Honda:2020txe} in the
continuum framework, may be regarded as a $U(1)$ analogue of the 't~Hooft
anomaly in the $SU(N)$ gauge theory studied in~Ref.~\cite{Gaiotto:2017yup}.

We make brief comments on some other related works. Realization of the
generalized symmetries on the (simplicial) lattice has already been studied in
detail in~Ref.~\cite{Kapustin:2014gua}. The fractional topological charge as a
function of the lattice gauge field is not considered
in~Ref.~\cite{Kapustin:2014gua}, however. In fact, in order to define a
topological charge in lattice gauge theory, a certain restriction on allowed
lattice gauge field configurations, such as
admissibility~\cite{Luscher:1981zq,Hernandez:1998et,Luscher:1998kn}, is
inevitable. Also the fractional topological charge in lattice gauge theory has
been studied over the years~\cite{Edwards:1998dj,deForcrand:2002vs,
Fodor:2009nh,Kitano:2017jng,Itou:2018wkm}. One possible method to obtain the
fractional topological charge associated with the $SU(N)/\mathbb{Z}_N$
principal bundle is to consider the gauge field in a higher representation as
blind for the $\mathbb{Z}_N$ part of the gauge transformation (such as the
adjoint representation) and divide the integer topological charge in that
higher representation by the corresponding Dynkin index ($2N$ for the adjoint
representation). In this method, however, one cannot explicitly specify the
underlying bundle structure such as the 't~Hooft flux or the $\mathbb{Z}_N$
two-form gauge field.

Presumably, the most analogous works to ours are
Refs.~\cite{Sulejmanpasic:2019ytl,Anosova:2022cjm}. Although
Refs.~\cite{Sulejmanpasic:2019ytl,Anosova:2022cjm} start with
\emph{non-compact\/} $U(1)$ link variables that are divided into disjoint
topological sectors (and thus the admissibility is implicit), the expression of
the lattice \emph{integer\/} topological charge is quite similar to ours.
In~Refs.~\cite{Sulejmanpasic:2019ytl,Anosova:2022cjm}, the Bianchi identity on
the $U(1)$ gauge field is further relaxed to introduce a static monopole and,
using the expression of the lattice topological charge, the Witten effect in
lattice gauge theory is observed. In the present paper, on the other hand, we
construct a fractional lattice topological charge in the compact $U(1)$
lattice gauge theory as a function of the 't~Hooft flux that is identified with
the gauge-invariant content of the $\mathbb{Z}_q$ two-form gauge field. It is
also natural in our construction to consider a static monopole by relaxing the
Bianchi identity; this point is left for future study. Also, the generalization
of our construction to the non-Abelian lattice gauge theory is an important
issue that we want to return in the near future.

\section{$U(1)/\mathbb{Z}_q$ principal bundle on~$T^4$ in $U(1)$ lattice
gauge theory}
\label{sec:2}
\subsection{Transition function}
\label{sec:2.1}
We consider a 4D periodic torus~$T^4$ of size~$L$; the Lorentz index is denoted
by $\mu$, $\nu$, \dots, etc.\ and runs over $1$, $2$, $3$, and~$4$:
\begin{equation}
   T^4\equiv\left\{
   x\in\mathbb{R}^4
   \mid\text{$0\leq x_\mu<L$ for all $\mu$}\right\}.
\label{eq:(2.1)}
\end{equation}
That is, any two points $x$ and~$y$ whose coordinates differ by an integer
multiple of~$L$ are identified, $x\sim y$.

We then consider a 4D lattice~$\Lambda$,
\begin{equation}
   \Lambda\equiv\left\{
   n\in\mathbb{Z}^4
   \mid\text{$0\leq n_\mu<L$ for all $\mu$}\right\},
\label{eq:(2.2)}
\end{equation}
by dividing $T^4$ into hypercubes~$c(n)$ specified by the lattice points
in~Eq.~\eqref{eq:(2.2)}:
\begin{equation}
   c(n)\equiv\left\{
   x\in\mathbb{R}^4
   \mid\text{$0\leq(x_\mu-n_\mu)\leq1$ for all $\mu$}\right\}.
\label{eq:(2.3)}
\end{equation}
We assume a $U(1)$ lattice gauge field on~$\Lambda$. The link variable
\begin{equation}
   U(n,\mu)\in U(1)
\label{eq:(2.4)}
\end{equation}
is residing on the link connecting $n$ and~$n+\Hat{\mu}$, where $\Hat{\mu}$
denotes a unit vector in the positive $\mu$~direction.

Following the idea of~Ref.~\cite{Luscher:1981zq}, we define the transition
function of the $U(1)/\mathbb{Z}_q$ on~$T^4$ by regarding each
hypercube~\eqref{eq:(2.3)} as the coordinate patch for~$T^4$. Thus, the
transition function is defined in the intersection between two hypercubes,
called the face:
\begin{equation}
   f(n,\mu)\equiv\left\{
   x\in c(n)
   \mid x_\mu=n_\mu\right\}
   =c(n-\Hat{\mu})\cap c(n).
\label{eq:(2.5)}
\end{equation}
This is a 3D cube. We then define the transition function at~$x\in f(n,\mu)$ by
\begin{equation}
   v_{n,\mu}(x)\equiv
   \omega_\mu(x)
   \Check{v}_{n,\mu}(x)\qquad
   \text{at $x\in f(n,\mu)$}.
\label{eq:(2.6)}
\end{equation}
In this expression, the first factor~$\omega_\mu(x)$ is given by
\begin{equation}
   \omega_\mu(x)
   \equiv
   \begin{cases}
   \exp\left(
   \frac{\pi i}{q}
   \sum_{\nu\neq\mu}\frac{z_{\mu\nu} x_\nu}{L}\right)
   &\text{for $x_\mu=0\bmod L$},
   \\
   1&\text{otherwise}.
   \end{cases}
\label{eq:(2.7)}
\end{equation}
Note that this is non-trivial only on the hyperplane $x_\mu=0\pmod L$. The
``twist angles'' $z_{\mu\nu}$ are integers and anti-symmetric in indices,
$z_{\mu\nu}=-z_{\nu\mu}$. Equation~\eqref{eq:(2.7)} is analogous to the loop
in~$SU(N)/\mathbb{Z}_N$ considered
in~Refs.~\cite{tHooft:1979rtg,vanBaal:1982ag}. In fact, along two non-trivial
intersecting one-cycles on~$T^4$, the factor~$\omega_\mu(x)$ defines a loop
in~$U(1)/\mathbb{Z}_q$. The winding of this loop gives rise to a fractional
topological charge~\cite{tHooft:1979rtg,vanBaal:1982ag}. The integers
$z_{\mu\nu}$ are the 't~Hooft flux or (the gauge-invariant content of) the
$\mathbb{Z}_q$ two-form gauge field; see below. $z_{\mu\nu}$ is defined only
modulo~$q$ and it labels elements of the cohomology
group~$H^2(T^4,\mathbb{Z}_q)$. In the present paper, these are fixed numbers
and non-dynamical.

The other factor~$\Check{v}_{n,\mu}(x)$ in~Eq.~\eqref{eq:(2.6)} is given by
L\"uscher's construction of the principal bundle in lattice gauge theory. When
the gauge group is~$U(1)$, the construction becomes very
simple~\cite{Fujiwara:2000wn} and it gives,
for~$x\in f(n,\mu)$ with~$\mu=1$, $2$, $3$, and~$4$, respectively,\footnote{In
deriving this, we have adopted the following definition of the standard
parallel transporter, which forms the basis of the construction
in~Ref.~\cite{Luscher:1981zq} (here $x\equiv n+\sum_{\mu=1}^4z_\mu\Hat{\mu}$):
\begin{equation}
   w^n(x)
   =U(n,4)^{z_4}U(n+z_4\Hat{4},3)^{z_3}
   U(n+z_4\Hat{4}+z_3\Hat{3},2)^{z_2}
   U(n+z_4\Hat{4}+z_3\Hat{3}+z_2\Hat{2},1)^{z_1}.
\label{eq:(2.8)}
\end{equation}}
\begin{align}
   \Check{v}_{n,1}(x)&=U(n-\Hat{1},1)
\notag\\
   &\qquad{}
   \times
   \exp\Bigl[
   iy_4\Check{F}_{14}(n-\Hat{1})
   +iy_3y_4\Check{F}_{13}(n-\Hat{1}+\Hat{4})
   +iy_3(1-y_4)\Check{F}_{13}(n-\Hat{1})
\notag\\
   &\qquad\qquad\qquad{}
   +iy_2y_3y_4\Check{F}_{12}(n-\Hat{1}+\Hat{3}+\Hat{4})
   +iy_2y_3(1-y_4)\Check{F}_{12}(n-\Hat{1}+\Hat{3})
\notag\\
   &\qquad\qquad\qquad{}
   +iy_2(1-y_3)y_4\Check{F}_{12}(n-\Hat{1}+\Hat{4})
   +iy_2(1-y_3)(1-y_4)\Check{F}_{12}(n-\Hat{1})
   \Bigr],
\notag\\
   \Check{v}_{n,2}(x)&=U(n-\Hat{2},2)
   \exp\left[
   iy_4\Check{F}_{24}(n-\Hat{2})
   +iy_3y_4\Check{F}_{23}(n-\Hat{2}+\Hat{4})
   +iy_3(1-y_4)\Check{F}_{23}(n-\Hat{2})
   \right],
\notag\\
   \Check{v}_{n,3}(x)&=U(n-\Hat{3},3)
   \exp\left[iy_4\Check{F}_{34}(n-\Hat{3})\right],
\notag\\
   \Check{v}_{n,4}(x)&=U(n-\Hat{4},4),
\label{eq:(2.9)}
\end{align}
where $y_\mu\equiv x_\mu-n_\mu$. The field strength in this expression is
defined by
\begin{equation}
   \Check{F}_{\mu\nu}(n)
   \equiv\frac{1}{iq}\ln
   \left[
   U(n,\mu)U(n+\Hat{\mu},\nu)U(n+\Hat{\nu},\mu)^{-1}U(n,\nu)^{-1}
   \right]^q\qquad
   -\pi<\Check{F}_{\mu\nu}(n)\leq\pi.
\label{eq:(2.10)}
\end{equation}
Here, the power~$q$ is supplemented inside the logarithm so that the field
strength is \emph{invariant\/} under the $\mathbb{Z}_q$ one-form gauge
transformation defined below.\footnote{Since $q$ is a positive integer,
Eq.~\eqref{eq:(2.10)} is equal to~$\ln[U(n,\mu)^qU(n+\Hat{\mu},\nu)^q%
U(n+\Hat{\nu},\mu)^{-q}U(n,\nu)^{-q}]/(iq)$; i.e., this is the logarithm of the
plaquette variable in the charge-$q$ representation.} We then require the
following admissibility condition,
\begin{equation}
   \sup_{n,\mu,\nu}\left|\Check{F}_{\mu\nu}(n)\right|<\epsilon\qquad
   0<\epsilon<\frac{\pi}{3q},
\label{eq:(2.11)}
\end{equation}
for allowed lattice gauge configurations. By applying the argument
in~Ref.~\cite{Luscher:1998kn} to~Eq.~\eqref{eq:(2.10)} carefully, one finds
that the condition $\epsilon<\pi/(3q)$ ensures that the Bianchi identity for
the field strength, i.e., the absence of the monopole current,
$\sum_{\nu,\rho,\sigma}%
\varepsilon_{\mu\nu\rho\sigma}\Delta_\nu\Check{F}_{\rho\sigma}(n)=0$,
holds.\footnote{Here and in what follows, the forward difference is defined
by~$\Delta_\mu f(n)\equiv f(n+\Hat{\mu})-f(n)$.}

\subsection{Cocycle condition}
\label{sec:2.2}
Let us examine the cocycle condition associated with the transition
function~\eqref{eq:(2.6)}. It is given by the product of the
transition functions in the intersection of four hypercubes, $c(n)$,
$c(n-\Hat{\mu})$, $c(n-\Hat{\nu})$, and~$c(n-\Hat{\mu}-\Hat{\nu})$, i.e.,
\begin{equation}
   p(n,\mu,\nu)\equiv\left\{
   x\in c(n)
   \mid x_\mu=n_\mu,x_\nu=n_\nu\right\}\qquad
   (\mu\neq\nu).
\label{eq:(2.12)}
\end{equation}
This is a 2D plaquette. Note that $p(n,\mu,\nu)$ is extending in directions
\emph{complementary\/} to $\mu$ and~$\nu$ in the present
convention~\cite{Luscher:1981zq}. When the gauge group is $U(1)/\mathbb{Z}_q$,
the cocycle can take a value in~$\mathbb{Z}_q$. Noting that the global
coordinate~$x_\mu$ on~$T^4$ possesses the discontinuity at~$x_\mu=0$,
for~$x\in p(n,\mu,\nu)$, we find
\begin{align}
   &v_{n-\Hat{\nu},\mu}(x)v_{n,\nu}(x)
   v_{n,\mu}(x)^{-1}v_{n-\Hat{\mu},\nu}(x)^{-1}
\notag\\
   &=\begin{cases}
   \exp\left(\frac{2\pi i}{q}z_{\mu\nu}\right)
   \in\mathbb{Z}_q&
   \text{for $x_\mu=x_\nu=0\bmod L$},
   \\
   1&\text{otherwise},
   \end{cases}
\label{eq:(2.13)}
\end{align}
where we have used the fact that the transition function~\eqref{eq:(2.9)}
satisfies the cocycle condition in the $U(1)$ gauge
theory~\cite{Luscher:1981zq}, $\Check{v}_{n-\Hat{\nu},\mu}(x)\Check{v}_{n,\nu}(x)
\Check{v}_{n,\mu}(x)^{-1}\Check{v}_{n-\Hat{\mu},\nu}(x)^{-1}=1$.\footnote{For the
expression in terms of the field strength in~Eq.~\eqref{eq:(2.9)} to fulfill
this cocycle condition, one has to use the Bianchi
identity~\cite{Fujiwara:2000wn}.} Thus the loop factor~$\omega_\mu(x)$
in~Eq.~\eqref{eq:(2.7)} gives rise to the ``$\mathbb{Z}_q$ breaking'' of the
cocycle condition.

\subsection{$\mathbb{Z}_q$ one-form global and gauge transformations}
\label{sec:2.3}
Let us now consider how the transition function transforms under the
$\mathbb{Z}_q$ one-form transformation. First, we identify the $\mathbb{Z}_q$
one-form \emph{global\/} transformation with the center transformation on link
variables crossing a 3D hypersurface, say~$n_\mu=0$. For instance, under
\begin{equation}
   U(n,\mu)\to\exp\left(\frac{2\pi i}{q}z_\mu\right)U(n,\mu)\qquad
   \text{$n_\mu=0$, $z_\mu\in\mathbb{Z}$ and
   $0\leq z_\mu<q$},
\label{eq:(2.14)}
\end{equation}
and $U(n,\mu)\to U(n,\mu)$ otherwise, the field strength~\eqref{eq:(2.10)} does
not change, the transition functions~\eqref{eq:(2.9)} are transformed as
\begin{equation}
   \Check{v}_{n,\mu}(x)
   \to
   \begin{cases}
   \exp\left(\frac{2\pi i}{q}z_\mu\right)\Check{v}_{n,\mu}(x)&
   \text{for $x_\mu=1$},
   \\
   \Check{v}_{n,\mu}(x)&\text{otherwise},
   \end{cases}
\label{eq:(2.15)}
\end{equation}
and the other transition functions $\Check{v}_{n,\nu\neq\mu}(x)$ do not change.
Since these are constant multiplications on the transition functions along the
hypersurface~$x_\mu=1$, the one-form global transformation does not affect the
cocycle condition,
$\Check{v}_{n-\Hat{\nu},\mu}(x)\Check{v}_{n,\nu}(x)\Check{v}_{n,\mu}^{-1}(x)
\Check{v}_{n-\Hat{\mu},\nu}^{-1}(x)=1$. Moreover, one can confirm that the
cocycle condition is not affected even under a smooth deformation of the
hypersurface. We thus conclude that the $\mathbb{Z}_q$ one-form global
transformation does not induce any further $\mathbb{Z}_q$ breaking of the
cocycle condition.

On the other hand, if we consider the $\mathbb{Z}_q$ one-form \emph{local\/}
or \emph{gauge\/} transformation defined by\footnote{Note that the field
strength~\eqref{eq:(2.10)} and the admissibility~\eqref{eq:(2.11)} are
invariant under this $\mathbb{Z}_q$ one-form gauge transformation.}
\begin{equation}
   U(n,\mu)\to\exp\left[\frac{2\pi i}{q}z_\mu(n)\right]U(n,\mu)\qquad
   \text{$z_\mu(n)\in\mathbb{Z}$, $0\leq z_\mu(n)<q$},
\label{eq:(2.16)}
\end{equation}
then the transition functions~\eqref{eq:(2.6)} are transformed as
\begin{equation}
   v_{n,\mu}(x)
   \to\exp\left[\frac{2\pi i}{q}z_\mu(n-\Hat{\mu})\right]
   v_{n,\mu}(x)\qquad
   x\in f(n,\mu).
\label{eq:(2.17)}
\end{equation}
The cocycle condition~\eqref{eq:(2.13)} is then modified accordingly to
\begin{equation}
   v_{n-\Hat{\nu},\mu}(x)v_{n,\nu}(x)
   v_{n,\mu}(x)^{-1}v_{n-\Hat{\mu},\nu}(x)^{-1}
   \equiv\exp\left[\frac{2\pi i}{q}z_{\mu\nu}(n-\Hat{\mu}-\Hat{\nu})\right],
\label{eq:(2.18)}
\end{equation}
where the $\mathbb{Z}_q$ two-form gauge field on the lattice is given by
\begin{equation}
   z_{\mu\nu}(n)
   =z_{\mu\nu}\delta_{n_\mu,L-1}\delta_{n_\nu,L-1}
   +\Delta_\mu z_\nu(n)
   -\Delta_\nu z_\mu(n)
   +qN_{\mu\nu}(n)\in\mathbb{Z}.
\label{eq:(2.19)}
\end{equation}
Thus the $\mathbb{Z}_q$ one-form gauge transformation~\eqref{eq:(2.16)} induces
an extra factor of a ``pure gauge'' form in the cocycle condition. Here, we
resolve the modulo~$q$ ambiguity of~$z_{\mu\nu}(n)$ by setting
\begin{equation}
   \begin{cases}
   0\leq z_{\mu\nu}(n)<q&\text{for~$\mu<\nu$},\\
   z_{\mu\nu}(n)\equiv-z_{\nu\mu}(n)&\text{for~$\mu>\nu$}.
   \end{cases}
\label{eq:(2.20)}
\end{equation}
The last lattice field $N_{\mu\nu}(n)\in\mathbb{Z}$ in~Eq.~\eqref{eq:(2.19)} is
required to restrict the value of~$z_{\mu\nu}(n)$ ($\mu<\nu$) in the
range~\eqref{eq:(2.20)}; recall that we have taken $0\leq z_\mu(n)<q$
in~Eq.~\eqref{eq:(2.16)}. Considering successive $\mathbb{Z}_q$ one-form gauge
transformations starting from~Eq.~\eqref{eq:(2.19)}, for a generic
$z_{\mu\nu}(n)$, we have
\begin{equation}
   z_{\mu\nu}(n)
   \to z_{\mu\nu}(n)+\Delta_\mu z_\nu(n)-\Delta_\nu z_\mu(n)
   +qN_{\mu\nu}(n).
\label{eq:(2.21)}
\end{equation}
(The fields $z_\mu(n)$ and~$N_{\mu\nu}(n)$ differ from those
in~Eq.~\eqref{eq:(2.19)}.) Our original configuration of the $\mathbb{Z}_q$
two-form gauge field in~Eq.~\eqref{eq:(2.13)},
$z_{\mu\nu}(n)=z_{\mu\nu}\delta_{n_\mu,L-1}\delta_{n_\nu,L-1}$, is flat, i.e.,
$(1/2)\sum_{\nu,\rho,\sigma}\varepsilon_{\mu\nu\rho\sigma}%
\Delta_\nu z_{\rho\sigma}(n)=0\bmod q$. This flatness of the $\mathbb{Z}_q$
two-form gauge field~\cite{Kapustin:2014gua,Tanizaki:2022ngt} is obviously
preserved under the $\mathbb{Z}_q$ one-form gauge transformation on the
lattice~\eqref{eq:(2.21)} because $N_{\mu\nu}(n)$ are integers and contribute to
the flatness condition only by an integer multiple of~$q$.
In~Appendix~\ref{sec:A}, we give a note on the flatness in our present lattice
formulation.

\section{Fractional topological charge}
\label{sec:3}
Now, the topological charge in the continuum,
\begin{equation}
   \mathcal{Q}=\frac{1}{32\pi^2}\int_{T^4}d^4x\,
   \varepsilon_{\mu\nu\rho\sigma}F_{\mu\nu}(x)F_{\rho\sigma}(x)
\label{eq:(3.1)}
\end{equation}
can be entirely expressed in terms of the transition function~$v_{n,\mu}(x)$ if
one divides $T^4$ into the cells~\eqref{eq:(2.3)} and uses the relation
between the gauge potentials in adjacent cells, $c(n-\Hat{\mu})$ and~$c(n)$. At
their overlap, $x\in f(n,\mu)$~\eqref{eq:(2.5)}, the relation is
\begin{equation}
   A_\lambda^{(n)}(x)
   =A_\lambda^{(n-\Hat{\mu})}(x)
   -iv_{n,\mu}(x)^{-1}\partial_\lambda v_{n,\mu}(x).
\label{eq:(3.2)}
\end{equation}
One then finds~\cite{Luscher:1981zq,vanBaal:1982ag}
\begin{equation}
   \mathcal{Q}=-\frac{1}{8\pi^2}\sum_{n\in\Lambda}
   \sum_{\mu,\nu,\rho,\sigma}\varepsilon_{\mu\nu\rho\sigma}\int_{p(n,\mu,\nu)}d^2x\,
   [v_{n,\mu}(x)\partial_\rho v_{n,\mu}(x)^{-1}]
   [v_{n-\Hat{\mu},\nu}(x)^{-1}\partial_\sigma v_{n-\Hat{\mu},\nu}(x)].
\label{eq:(3.3)}
\end{equation}
As noted in~Ref.~\cite{vanBaal:1982ag}, this expression holds even if the
cocycle condition is relaxed by $\mathbb{Z}_q$ elements
as~Eq.~\eqref{eq:(2.13)}. Note that this expression is manifestly invariant
under the $\mathbb{Z}_q$ one-form gauge transformation~\eqref{eq:(2.17)}
because the extra factor in~Eq.~\eqref{eq:(2.17)} is independent of the
continuous coordinate~$x$.

We thus substitute our transition function~\eqref{eq:(2.6)}
into~Eq.~\eqref{eq:(3.3)}. After some calculation using the Bianchi identity,
we have
\begin{align}
   \mathcal{Q}&=
   \frac{1}{8q^2}\sum_{\mu,\nu,\rho,\sigma}\varepsilon_{\mu\nu\rho\sigma}
   z_{\mu\nu}z_{\rho\sigma}
   +\frac{1}{8\pi q}\sum_{\mu,\nu,\rho,\sigma}\varepsilon_{\mu\nu\rho\sigma}
   z_{\mu\nu}\sum_{n\in\Lambda,n_\mu=0}\Check{F}_{\rho\sigma}(n)
\notag\\
   &\qquad{}
   +\frac{1}{32\pi^2}\sum_{n\in\Lambda}\sum_{\mu,\nu,\rho,\sigma}
   \varepsilon_{\mu\nu\rho\sigma}
   \Check{F}_{\mu\nu}(n)
   \Check{F}_{\rho\sigma}(n+\Hat{\mu}+\Hat{\nu}).
\label{eq:(3.4)}
\end{align}
On the right-hand side, the last term is the well-known expression of the
topological charge in the $U(1)$ lattice gauge
theory~\cite{Luscher:1998kn,Fujiwara:1999fi,Fujiwara:2000wn}. It takes integer
values for admissible gauge fields and thus is topological. The first term on
the right-hand side gives a fractional topological charge associated with the
't~Hooft flux (the winding of a non-trivial cycle to
$U(1)/\mathbb{Z}_q$)~\cite{vanBaal:1982ag}. The second term is a ``cross term''
and it sums the first Chern numbers on~$T^2$ in the $\rho\sigma$ direction
over~$n_\nu$. Under the admissibility, the first Chern number is also quantized
on the lattice (see Ref.~\cite{Fujiwara:2000wn}) and
\begin{equation}
   \sum_{n\in\Lambda,n_\mu=0,n_\nu=\text{fixed}}
   \sum_{\rho,\sigma}\varepsilon_{\mu\nu\rho\sigma}
   \Check{F}_{\rho\sigma}(n)
   =4\pi\mathbb{Z}.
\label{eq:(3.5)}
\end{equation}
By construction, the lattice topological charge~\eqref{eq:(3.4)} is invariant
under the $\mathbb{Z}_q$ one-form \emph{gauge\/} transformation
in~Eq.~\eqref{eq:(2.16)}.

The lattice topological charge~\eqref{eq:(3.4)} also possesses a simple
transformation property under the time reversal. We may define the time
reversal transformation~$\mathcal{T}$ on a lattice field by
\begin{equation}
   U(n,\mu)\stackrel{\mathcal{T}}{\to}
   \begin{cases}
   U(\Bar{n},\mu)&\text{for $\mu\neq4$},\\
   U(\Bar{n}-\Hat{4},4)^{-1}&\text{for $\mu=4$},\\
   \end{cases}
\label{eq:(3.6)}
\end{equation}
where $\Bar{n}\equiv(n_1,n_2,n_3,-n_4)$. Under this, the field
strength~\eqref{eq:(2.10)} is transformed as
\begin{equation}
   \Check{F}_{\mu\nu}(n)\stackrel{\mathcal{T}}{\to}
   \begin{cases}
   \Check{F}_{\mu\nu}(\Bar{n})&\text{for $\mu\neq4$, $\nu\neq4$},\\
   -\Check{F}_{4\nu}(\Bar{n}-\Hat{4})&\text{for $\mu=4$},\\
   -\Check{F}_{\mu4}(\Bar{n}-\Hat{4})&\text{for $\nu=4$}.\\
   \end{cases}
\label{eq:(3.7)}
\end{equation}
Note that this transformation preserves the admissibility~\eqref{eq:(2.11)}.
Using these and the Bianchi identity, it can be seen that the topological
charge~\eqref{eq:(3.4)} changes its sign under the time reversal
transformation,
\begin{equation}
   \mathcal{Q}\stackrel{\mathcal{T}}{\to}-\mathcal{Q},
\label{eq:(3.8)}
\end{equation}
if we do the time reversal transformation on the 't~Hooft flux at the same
time:
\begin{equation}
   z_{\mu\nu}
   \stackrel{\mathcal{T}}{\to}\begin{cases}
   z_{\mu\nu}&\text{for $\mu\neq4$, $\nu\neq4$},\\
   -z_{4\nu}&\text{for $\mu=4$},\\
   -z_{\mu4}&\text{for $\nu=4$}.\\
   \end{cases}
\label{eq:(3.9)}
\end{equation}
This transformation may be generalized to the time reversal of the two-form
gauge field as
\begin{equation}
   z_{\mu\nu}(n)
   \stackrel{\mathcal{T}}{\to}\begin{cases}
   z_{\mu\nu}(\Bar{n})&\text{for $\mu\neq4$, $\nu\neq4$},\\
   -z_{4\nu}(\Bar{n}+\Hat{4})&\text{for $\mu=4$},\\
   -z_{\mu4}(\Bar{n}+\Hat{4})&\text{for $\nu=4$},\\
   \end{cases}
\label{eq:(3.10)}
\end{equation}
so that this is consistent with~Eq.~\eqref{eq:(2.19)}.

Now the 't~Hooft flux in the topological charge~\eqref{eq:(3.4)} is constant.
However, we may rewrite this expression by using the \emph{local\/}
$\mathbb{Z}_q$ two-form gauge field~$z_{\mu\nu}(n)$. Recalling
Eq.~\eqref{eq:(2.19)}, we have
\begin{equation}
   \mathcal{Q}=
   \frac{1}{32\pi^2}\sum_{n\in\Lambda}\sum_{\mu,\nu,\rho,\sigma}
   \varepsilon_{\mu\nu\rho\sigma}
   \left[
   F_{\mu\nu}(n)
   +\frac{2\pi}{q}z_{\mu\nu}(n)\right]
   \left[
   F_{\rho\sigma}(n+\Hat{\mu}+\Hat{\nu})
   +\frac{2\pi}{q}z_{\rho\sigma}(n+\Hat{\mu}+\Hat{\nu})\right],
\label{eq:(3.11)}
\end{equation}
where we have introduced another field strength on the lattice by
\begin{equation}
   F_{\mu\nu}(n)
   \equiv \Check{F}_{\mu\nu}(n)
   -\frac{2\pi}{q}
   \left[
   \Delta_\mu z_\nu(n)-\Delta_\nu z_\mu(n)+qN_{\mu\nu}(n)
   \right]
\label{eq:(3.12)}
\end{equation}
using the functions appearing in~Eq.~\eqref{eq:(2.19)}. Under the
$\mathbb{Z}_q$ one-form gauge transformation~\eqref{eq:(2.21)}, the new
field strength is not invariant and transforms as
\begin{equation}
   F_{\mu\nu}(n)
   \to F_{\mu\nu}(n)
   -\frac{2\pi}{q}
   \left[
   \Delta_\mu z_\nu(n)-\Delta_\nu z_\mu(n)+qN_{\mu\nu}(n)
   \right].
\label{eq:(3.13)}
\end{equation}
Compared with~Eq.~\eqref{eq:(3.4)}, the locality of the topological
charge~$\mathcal{Q}$ is manifest with the
expression~\eqref{eq:(3.11)}.\footnote{Note that the integer field
$N_{\mu\nu}(n)$ in~Eq.~\eqref{eq:(2.19)} is determined from~$z_\mu(n)$ locally.}

\section{'t~Hooft anomaly}
\label{sec:4}
\subsection{General setting}
\label{sec:4.1}
Our construction above may be employed to consider, with lattice
regularization, the mixed 't~Hooft anomaly between the $\mathbb{Z}_q$ one-form
symmetry and the time reversal symmetry in the $U(1)$ gauge theory with matter
fields of charge~$q\in2\mathbb{Z}$, when the vacuum angle~$\theta$
is~$\pi$~\cite{Honda:2020txe}. We can assume a lattice action that is invariant
under the $\mathbb{Z}_q$ one-form global transformation:\footnote{To
incorporate the admissibility and the smoothness of the lattice action, a more
ingenious construction such as the one in~Ref.~\cite{Luscher:1998du} would be
desirable; this point is irrelevant in the present discussion concerning
symmetries of the action.}
\begin{equation}
   S\equiv\frac{1}{4g_0^2}\sum_{n\in\Lambda}\sum_{\mu,\nu}
   \Check{F}_{\mu\nu}(n)\Check{F}_{\mu\nu}(n)
   +S_{\text{matter}}
   -iq\theta\mathcal{Q},
\label{eq:(4.1)}
\end{equation}
where $g_0$ is the bare coupling. This original system does not contain the
$\mathbb{Z}_q$ two-form gauge field. We also assume that the lattice action
except the last topological term is even under the time reversal
transformation; this is actually the case for the first pure gauge term
in~Eq.~\eqref{eq:(4.1)}.

In the last topological term in~Eq.~\eqref{eq:(4.1)}, $\theta$ is multiplied
by~$q$ as~$iq\theta\mathcal{Q}$ instead of~$i\theta\mathcal{Q}$. This is
because, with the conventional normalization of~$\theta$, the Witten
effect~\cite{Witten:1979ey} suggests a $2\pi q$~periodicity of~$\theta$
instead of the $2\pi$~periodicity. This rescaling of the periodicity actually
occurs at least in the Cardy--Rabinovici
model~\cite{Cardy:1981qy,Cardy:1981fd} as studied in~Ref.~\cite{Honda:2020txe}
(see also~Ref.~\cite{Hidaka:2019jtv}). That is, the full spectrum of the system
including the monopole and dyons is invariant only under a $2\pi q$ shift of
the original vacuum angle, instead of a $2\pi$ shift; thus the periodicity
of~$\theta$ is~$2\pi$ only with the combination in~Eq.~\eqref{eq:(4.1)}. Since
we do not introduce the monopole and dyons in the present lattice setup, we
cannot observe the Witten effect and the associated rescaling of the vacuum
angle directly. Nevertheless, it is interesting to see a possible 't~Hooft
anomaly in our lattice regularized setup, temporarily accepting the above
rescaling of the vacuum angle. This is what we do here.

Now, let us set $\theta=\pi$ in~Eq.~\eqref{eq:(4.1)}. The original partition
function is then time reversal invariant because
$i\pi q\mathcal{Q}\stackrel{\mathcal{T}}{\to}-i\pi q\mathcal{Q}
\sim+i\pi q\mathcal{Q}$ because of the postulated $2\pi$ periodicity
of~$\theta$ (the original $\mathcal{Q}$ is an integer).

We then switch the $\mathbb{Z}_q$ two-form gauge field on.
From~Eqs.~\eqref{eq:(3.4)} and~\eqref{eq:(3.5)}, we have
\begin{equation}
   q\mathcal{Q}=
   \frac{1}{8q}\sum_{\mu,\nu,\rho,\sigma}\varepsilon_{\mu\nu\rho\sigma}
   z_{\mu\nu}z_{\rho\sigma}
   +\mathbb{Z}.
\label{eq:(4.2)}
\end{equation}
The factor~$e^{i\pi q\mathcal{Q}}$ in the integrand of the functional integral
then acquires an extra phase factor under the time reversal, as (recall
Eq.~\eqref{eq:(3.8)})
\begin{align}
   e^{i\pi q\mathcal{Q}}
   &\stackrel{\mathcal{T}}{\to}
   e^{-i\pi q\mathcal{Q}}
   =e^{-2\pi iq\mathcal{Q}}\cdot e^{i\pi q\mathcal{Q}}
\notag\\
   &=\exp\left(
   -\frac{2\pi i}{8q}\sum_{\mu,\nu,\rho,\sigma}\varepsilon_{\mu\nu\rho\sigma}
   z_{\mu\nu}z_{\rho\sigma}
   \right)e^{i\pi q\mathcal{Q}}.
\label{eq:(4.3)}
\end{align}

\subsection{Construction of the local counterterm}
\label{sec:4.2}
One should then ask~\cite{Gaiotto:2017yup} whether a certain local
gauge-invariant term of the $\mathbb{Z}_q$ two-form gauge field~$z_{\mu\nu}(n)$
can counter the above breaking of the time reversal symmetry. Using the
technique reviewed in~Appendix~\ref{sec:B}, it can be seen that a local term
that transforms ``covariantly'' under the lattice $\mathbb{Z}_q$ one-form gauge
transformation~\eqref{eq:(2.21)} is given by
\begin{equation}
   \exp\left[
   \frac{2\pi ik}{4q}\sum_{n\in\Lambda}\sum_{\mu,\nu,\rho,\sigma}
   \varepsilon_{\mu\nu\rho\sigma}
   z_{\mu\nu}(n)z_{\rho\sigma}(n+\Hat{\mu}+\Hat{\nu})
   \right].
\label{eq:(4.4)}
\end{equation}
It is easy to see that, under the $\mathbb{Z}_q$ one-form gauge
transformation~\eqref{eq:(2.21)}, the combination
$\sum_{n\in\Lambda}\sum_{\mu,\nu,\rho,\sigma}\varepsilon_{\mu\nu\rho\sigma}
z_{\mu\nu}(n)z_{\rho\sigma}(n+\Hat{\mu}+\Hat{\nu})$ shifts
by~$4q\mathbb{Z}$; see Appendix~\ref{sec:B}. Therefore, for the phase
factor~\eqref{eq:(4.4)} to be gauge invariant, the constant~$k$ must be an
integer, $k\in\mathbb{Z}$.

However, Eq.~\eqref{eq:(4.4)}, which would be regarded as the wedge product of
the elements of~$H^2(T^4,\mathbb{Z}_q)$ on the hypercubic lattice, is not the
counterterm with the ``finest'' coefficient, as known for the corresponding
counterterm on the simplicial
lattice~\cite{Kapustin:2013qsa,Kapustin:2014gua}.\footnote{We owe the following
discussion to Yuya Tanizaki.} The counterterm with a ``finer'' coefficient
on~$T^4$ can be constructed by employing the integral lift
of~$H^2(T^4,\mathbb{Z}_q)$
to~$H^2(T^4,\mathbb{Z})$~\cite{Kapustin:2013qsa,Kapustin:2014gua}. On our
periodic hypercubic lattice~$\Lambda$, we may construct an analogue of the
integral lift, $\Bar{z}_{\mu\nu}(n)$, which satisfies the flatness,
$(1/2)\sum_{\nu,\rho,\sigma}\varepsilon_{\mu\nu\rho\sigma}\Delta_\nu
\Bar{z}_{\rho\sigma}(n)=0$ (strictly zero \emph{not\/} modulo~$q$),
from~$z_{\mu\nu}(n)$ as follows.

First, we define an integer
$m_\mu(n)\equiv(1/2)\sum_{\nu,\rho,\sigma}\varepsilon_{\mu\nu\rho\sigma}
\Delta_\nu z_{\rho\sigma}(n)/q$ for each 3D cube, which spans from the site~$n$
to directions complementary to~$\mu$. $m_\mu(n)$ is an integer, because of the
modulo~$q$ flatness of~$z_{\mu\nu}(n)$,
$(1/2)\sum_{\nu,\rho,\sigma}\varepsilon_{\mu\nu\rho\sigma}\Delta_\nu
z_{\rho\sigma}(n)\in q\mathbb{Z}$.

Next, we take a 3D space specified by a fixed~$n_\mu$ and consider paths
connecting centers of cubes in the 3D space. The paths are defined such that
$|m_\mu(n)|$ paths begin from the cube at~$n$ if $m_\mu(n)>0$, while
$|m_\mu(n)|$ paths end at the cube if $m_\mu(n)<0$. A certain consistent
configuration of paths can be defined in this way, because the total sum
of~$m_\mu(n)$ on the three-dimensional space identically vanishes,
$\sum_{n\in\Lambda,\text{$n_\mu$ fixed}}m_\mu(n)
=\sum_{n\in\Lambda,\text{$n_\mu$ fixed}}
(1/2)\sum_{\nu,\rho,\sigma}\varepsilon_{\mu\nu\rho\sigma}
\Delta_\nu z_{\rho\sigma}(n)/q=0$. Then, it is obvious that $\Bar{z}_{\mu\nu}(n)$
such that $\sum_{\nu,\rho,\sigma}\varepsilon_{\mu\nu\rho\sigma}\Delta_\nu
\Bar{z}_{\rho\sigma}(n)=0$ can be obtained by subtracting $qm_{\mu\nu}(n)$
from~$z_{\mu\nu}(n)$. Here, $m_{\mu\nu}(n)$ is the signed number of paths going
through the plaquette at which $z_{\mu\nu}(n)$ is residing; the sign
of~$m_{\mu\nu}(n)$ is defined to be positive if the paths go through the
plaquette in the direction of the standard orientation of the plaquette and
negative if they go through in the opposite direction. This construction gives
the relation $\Bar{z}_{\mu\nu}(n)=z_{\mu\nu}(n)-qm_{\mu\nu}(n)$, where
$m_{\nu\mu}(n)=-m_{\mu\nu}(n)$.

The integral lift $\Bar{z}_{\mu\nu}(n)$ can differ depending on the choice of
the configuration of paths but the difference can be expressed in the
difference of the field~$m_{\mu\nu}(n)$. Since
$\Bar{z}_{\mu\nu}(n)=z_{\mu\nu}(n)-qm_{\mu\nu}(n)$ has the form of the gauge
transformation~\eqref{eq:(2.21)}, the choice of the configuration of paths does
not matter as far as gauge-invariant quantities (such as the counterterm
below) are concerned. Also, it is obvious from~Eq.~\eqref{eq:(2.21)} that the
gauge transformation of~$\Bar{z}_{\mu\nu}(n)$ takes the form
$\Bar{z}_{\mu\nu}(n)\to\Bar{z}_{\mu\nu}(n)+\Delta_\mu z_\nu(n)-\Delta_\nu z_\mu(n)
+q\Bar{N}_{\mu\nu}(n)$, where $\Bar{N}_{\mu\nu}(n)\in\mathbb{Z}$. Because of the
flatness of~$\Bar{z}_{\mu\nu}(n)$, we thus infer that
$\Bar{N}_{\mu\nu}(n)$ is also flat,
$\sum_{\nu,\rho,\sigma}\varepsilon_{\mu\nu\rho\sigma}\Delta_\nu
\Bar{N}_{\rho\sigma}(n)=0$.

Finally, when $\sum_{\nu,\rho,\sigma}\varepsilon_{\mu\nu\rho\sigma}\Delta_\nu
\Bar{z}_{\rho\sigma}(n)
=\sum_{\nu,\rho,\sigma}\varepsilon_{\mu\nu\rho\sigma}\Delta_\nu
\Bar{N}_{\rho\sigma}(n)=0$, which can be expressed as $d\Bar{z}^{(2)}=d\Bar{N}=0$
in the notation of Appendix~\ref{sec:B}, by employing the argument
in~Ref.~\cite{Fujiwara:2000wn}, one can see that the combination
$\sum_{n\in\Lambda}\sum_{\mu,\nu,\rho,\sigma}\varepsilon_{\mu\nu\rho\sigma}
\Bar{z}_{\mu\nu}(n)\Bar{z}_{\rho\sigma}(n+\Hat{\mu}+\Hat{\nu})$
shifts by~$8q\mathbb{Z}$ (\emph{not\/} $4q\mathbb{Z}$) under the gauge
transformation.

Therefore, the counterterm with a finer coefficient is given by
\begin{equation}
   e^{-S_{\text{counter}}}
   =\exp\left[
   \frac{2\pi ik}{8q}\sum_{n\in\Lambda}\sum_{\mu,\nu,\rho,\sigma}
   \varepsilon_{\mu\nu\rho\sigma}
   \Bar{z}_{\mu\nu}(n)\Bar{z}_{\rho\sigma}(n+\Hat{\mu}+\Hat{\nu})
   \right]
\label{eq:(4.5)}
\end{equation}
and this is gauge-invariant for~$k\in\mathbb{Z}$ (note the difference in
coefficients in~Eqs.~\eqref{eq:(4.4)} and~\eqref{eq:(4.5)}). We expect that
this is the finest gauge invariant coefficient from the corresponding result in
the continuum theory~\cite{Honda:2020txe}.

Since our representative configuration in~Eq.~\eqref{eq:(2.13)},
$z_{\mu\nu}(n)=z_{\mu\nu}\delta_{n_\mu,L-1}\delta_{n_\nu,L-1}$, is flat,
$\sum_{\nu,\rho,\sigma}\varepsilon_{\mu\nu\rho\sigma}%
\Delta_\nu z_{\rho\sigma}(n)=0$, the corresponding integral lift is also given by
this, $\Bar{z}_{\mu\nu}(n)=z_{\mu\nu}\delta_{n_\mu,L-1}\delta_{n_\nu,L-1}$.
Substituting this into~Eq.~\eqref{eq:(4.5)} yields
\begin{equation}
   e^{-S_{\text{counter}}}
   =\exp\left(
   \frac{2\pi ik}{8q}\sum_{\mu,\nu,\rho,\sigma}
   \varepsilon_{\mu\nu\rho\sigma}
   z_{\mu\nu}z_{\rho\sigma}
   \right).
\label{eq:(4.6)}
\end{equation}
Therefore, after the addition of the counterterm in~Eq.~\eqref{eq:(4.5)},
Eq.~\eqref{eq:(4.3)} is modified to
\begin{align}
   e^{i\pi q\mathcal{Q}}e^{-S_{\text{counter}}}
   &\stackrel{\mathcal{T}}{\to}
   e^{-i\pi q\mathcal{Q}}
   e^{+S_{\text{counter}}}
   =e^{-2i\pi q\mathcal{Q}}e^{2S_{\text{counter}}}
   e^{i\pi q\mathcal{Q}}e^{-S_{\text{counter}}}
\notag\\
   &=
   \exp\left[
   -\frac{2\pi i(2k+1)}{8q}
   \sum_{\mu,\nu,\rho,\sigma}
   \varepsilon_{\mu\nu\rho\sigma}
   z_{\mu\nu}z_{\rho\sigma}
   \right]
   e^{i\pi q\mathcal{Q}}e^{-S_{\text{counter}}}.
\label{eq:(4.7)}
\end{align}
Since the possible minimal non-zero value
of~$|\sum_{\mu,\nu,\rho,\sigma}\varepsilon_{\mu\nu\rho\sigma}z_{\mu\nu}z_{\rho\sigma}|$
is~$8$ (for instance, the choice, $z_{12}=z_{34}=1$ and other components
vanish, gives this), this shows that if we can choose the integer~$k$ such that
$2k+1=0\bmod q$, the anomaly is countered. This is impossible for even $q$ and
possible for odd $q$. This thus implies the mixed 't~Hooft anomaly between the
$\mathbb{Z}_q$ one-form symmetry and the time reversal symmetry
for~$q\in2\mathbb{Z}$ when~$\theta=\pi$~\cite{Honda:2020txe}.

\section{Conclusion}
\label{sec:5}
In this paper, assuming an appropriate admissibility condition on allowed
lattice field configurations, we constructed the transition function of the
$U(1)/\mathbb{Z}_q$ principal bundle on~$T^4$ from the compact $U(1)$ lattice
gauge field by combining L\"uscher's method and a loop factor
in~$U(1)/\mathbb{Z}_q$. The resulting topological charge takes fractional
values and is invariant under the $\mathbb{Z}_q$ one-form gauge transformation.
Also, the topological charge is odd under the lattice time reversal
transformation. From these properties, assuming a rescaling of the vacuum angle
$\theta\to q\theta$ suggested by the Witten effect, our construction provides a
lattice implementation of the mixed 't~Hooft anomaly between the
$\mathbb{Z}_q$ one-form symmetry and the time reversal symmetry in the $U(1)$
gauge theory with matter fields of charge~$q\in2\mathbb{Z}$ when
$\theta=\pi$~\cite{Honda:2020txe}. This may be regarded as a $U(1)$ analogue of
the mixed 't~Hooft anomaly between the $\mathbb{Z}_N$ one-form symmetry and the
time reversal symmetry in the $SU(N)$ gauge theory with~$N\in2\mathbb{Z}$
with~$\theta=\pi$~\cite{Gaiotto:2017yup}. For odd~$q>1$, which requires
$k\neq0$ in~Eq.~\eqref{eq:(4.5)} for the anomaly cancellation at~$\theta=\pi$,
we may consider a global inconsistency between different values of~$\theta$,
imitating the discussions in~Refs.~\cite{Gaiotto:2017yup,Tanizaki:2017bam}.

Although our construction of the transition function and the fractional
topological charge is perfectly legitimate, our discussion on the mixed
't~Hooft anomaly is still incomplete because we have simply assumed the
rescaling of the vacuum angle without introducing the monopole and dyons. To
observe the Witten effect, these degrees of freedom should be incorporated into
our treatment. For this, we have to relax the Bianchi identity and it appears
that the works~\cite{Sulejmanpasic:2019ytl,Anosova:2022cjm} are quite
suggestive in this aspect.

Generalization of our construction to non-Abelian lattice gauge theory is an
important issue that we want to return to in the near future.

\section*{Acknowledgements}
We would like to thank Yoshimasa Hidaka, Satoshi Yamaguchi, and especially Yuya
Tanizaki for helpful discussions.
We also thank Yuki Miyakawa and Soma Onoda for collaboration.
This work was partially supported by Japan Society for the Promotion of Science
(JSPS) Grant-in-Aid for Scientific Research Grant Numbers JP21J30003~(O.M.)
and~JP20H01903~(H.S.).

\appendix

\section{Flatness of the $\mathbb{Z}_q$ two-form gauge field}
\label{sec:A}
The flatness of the $\mathbb{Z}_q$ two-form gauge field follows from the
consistency of transition functions among ``quadruple''
overlap~\cite{Kapustin:2014gua,Tanizaki:2022ngt} and, for our square lattice,
it may be seen in the following way.\footnote{The argument in this appendix
holds even in non-Abelian lattice gauge theory.}

We take a point~$x$ on the link connecting $n$ and~$n+\Hat{4}$ and consider
transitions among the following eight hypercubes, which share the above link:
\begin{align}
   &c(n)\qquad
   c(n-\Hat{1})\qquad
   c(n-\Hat{2})\qquad
   c(n-\Hat{3})
\notag\\
   &c(n-\Hat{1}-\Hat{2})\qquad
   c(n-\Hat{1}-\Hat{3})\qquad
   c(n-\Hat{2}-\Hat{3})\qquad
   c(n-\Hat{1}-\Hat{2}-\Hat{3}).
\label{eq:(A1)}
\end{align}
We start from the combination
\begin{align}
   &v_{\mathstrut\smash{n-\Hat{3},2}}(x)
   v_{\mathstrut\smash{n,3}}(x)
   v_{\mathstrut\smash{n,2}}(x)^{-1}
   v_{\mathstrut\smash{n-\Hat{2},3}}(x)^{-1}
   \times
   v_{\mathstrut\smash{n-\Hat{1}-\Hat{2},3}}(x)
   v_{\mathstrut\smash{n-\Hat{1},2}}(x)
   v_{\mathstrut\smash{n-\Hat{1},3}}(x)^{-1}
   v_{\mathstrut\smash{n-\Hat{1}-\Hat{3},2}}(x)^{-1}
\notag\\
   &=\exp\left[
   \frac{2\pi i}{q}\Delta_1z_{23}(n-\Hat{1}-\Hat{2}-\Hat{3})
   \right],
\label{eq:(A2)}
\end{align}
where we have used the relation~\eqref{eq:(2.18)}. Since the right-hand side
of~Eq.~\eqref{eq:(2.18)} is an element of~$\mathbb{Z}_q$, the left-hand side is
invariant under any similarity transformation. Using this fact, we can rewrite
Eq.~\eqref{eq:(A2)} as
\begin{align}
   &v_{\mathstrut\smash{n-\Hat{1},3}}(x)^{-1}
   v_{\mathstrut\smash{n-\Hat{1}-\Hat{3},2}}(x)^{-1}
\notag\\
   &\qquad{}
   \times
   v_{\mathstrut\smash{n-\Hat{2}-\Hat{3},1}}(x)
\notag\\
   &\qquad{}
   \times
   v_{\mathstrut\smash{n-\Hat{3},2}}(x)
   v_{\mathstrut\smash{n,3}}(x)
   \left[v_{\mathstrut\smash{n,1}}(x)^{-1}
   v_{\mathstrut\smash{n,1}}(x)\right]
   v_{\mathstrut\smash{n,2}}(x)^{-1}
   v_{\mathstrut\smash{n-\Hat{2},3}}(x)^{-1}
\notag\\
   &\qquad{}
   \times
   v_{\mathstrut\smash{n-\Hat{2}-\Hat{3},1}}(x)^{-1}
\notag\\
   &\qquad{}
   \times
   v_{\mathstrut\smash{n-\Hat{1}-\Hat{2},3}}(x)
   v_{\mathstrut\smash{n-\Hat{1},2}}(x)
   v_{\mathstrut\smash{n-\Hat{1},3}}(x)^{-1}
   v_{\mathstrut\smash{n-\Hat{1}-\Hat{3},2}}(x)^{-1}
\notag\\
   &\qquad{}
   \times
   \left[
   v_{\mathstrut\smash{n-\Hat{1},3}}(x)^{-1}
   v_{\mathstrut\smash{n-\Hat{1}-\Hat{3},2}}(x)^{-1}
   \right]^{-1}
\notag\\
   &=v_{\mathstrut\smash{n-\Hat{1},3}}(x)^{-1}
   v_{\mathstrut\smash{n-\Hat{1}-\Hat{3},2}}(x)^{-1}
   v_{\mathstrut\smash{n-\Hat{2}-\Hat{3},1}}(x)
   v_{\mathstrut\smash{n-\Hat{3},2}}(x)
   v_{\mathstrut\smash{n,3}}(x)
   v_{\mathstrut\smash{n,1}}(x)^{-1}
\notag\\
   &\qquad{}
   \times
   v_{\mathstrut\smash{n,1}}(x)
   v_{\mathstrut\smash{n,2}}(x)^{-1}
   v_{\mathstrut\smash{n-\Hat{2},3}}(x)^{-1}
   v_{\mathstrut\smash{n-\Hat{2}-\Hat{3},1}}(x)^{-1}
   v_{\mathstrut\smash{n-\Hat{1}-\Hat{2},3}}(x)
   v_{\mathstrut\smash{n-\Hat{1},2}}(x).
\label{eq:(A3)}
\end{align}
The factor on the right-hand side can be written as
\begin{align}
   &v_{\mathstrut\smash{n-\Hat{1},3}}(x)^{-1}
   v_{\mathstrut\smash{n-\Hat{1}-\Hat{3},2}}(x)^{-1}
   v_{\mathstrut\smash{n-\Hat{2}-\Hat{3},1}}(x)
   v_{\mathstrut\smash{n-\Hat{3},2}}(x)
   v_{\mathstrut\smash{n,3}}(x)
   v_{\mathstrut\smash{n,1}}(x)^{-1}
\notag\\
   &=v_{\mathstrut\smash{n-\Hat{1},3}}(x)^{-1}
   v_{\mathstrut\smash{n-\Hat{1}-\Hat{3},2}}(x)^{-1}
   v_{\mathstrut\smash{n-\Hat{2}-\Hat{3},1}}(x)
   v_{\mathstrut\smash{n-\Hat{3},2}}(x)
\notag\\
   &\qquad{}
   \times
   v_{\mathstrut\smash{n-\Hat{3},1}}(x)^{-1}v_{\mathstrut\smash{n-\Hat{1},3}}(x)
   \left[
   v_{\mathstrut\smash{n-\Hat{3},1}}(x)^{-1}v_{\mathstrut\smash{n-\Hat{1},3}}(x)
   \right]^{-1}
   v_{\mathstrut\smash{n,3}}(x)
   v_{\mathstrut\smash{n,1}}(x)^{-1}
\notag\\
   &=v_{\mathstrut\smash{n-\Hat{1},3}}(x)^{-1}\times
   v_{\mathstrut\smash{n-\Hat{1}-\Hat{3},2}}(x)^{-1}
   v_{\mathstrut\smash{n-\Hat{2}-\Hat{3},1}}(x)
   v_{\mathstrut\smash{n-\Hat{3},2}}(x)
   v_{\mathstrut\smash{n-\Hat{3},1}}(x)^{-1}
   \times
   v_{\mathstrut\smash{n-\Hat{1},3}}(x)
\notag\\
   &\qquad{}
   \times
   v_{\mathstrut\smash{n-\Hat{1},3}}(x)^{-1}
   v_{\mathstrut\smash{n-\Hat{3},1}}(x)
   v_{\mathstrut\smash{n,3}}(x)
   v_{\mathstrut\smash{n,1}}(x)^{-1}
\notag\\
   &=\exp\left\{
   \frac{2\pi i}{q}
   \left[-z_{31}(n-\Hat{1}-\Hat{3})+z_{12}(n-\Hat{1}-\Hat{2}-\Hat{3})\right]
   \right\}.
\label{eq:(A4)}
\end{align}
In a similar way, we find
\begin{align}
   &v_{\mathstrut\smash{n,1}}(x)
   v_{\mathstrut\smash{n,2}}(x)^{-1}
   v_{\mathstrut\smash{n-\Hat{2},3}}(x)^{-1}
   v_{\mathstrut\smash{n-\Hat{2}-\Hat{3},1}}(x)^{-1}
   v_{\mathstrut\smash{n-\Hat{1}-\Hat{2},3}}(x)
   v_{\mathstrut\smash{n-\Hat{1},2}}(x).
\notag\\
   &=\exp\left\{
   \frac{2\pi i}{q}
   \left[z_{31}(n-\Hat{1}-\Hat{2}-\Hat{3})-z_{12}(n-\Hat{1}-\Hat{2})\right]
   \right\}.
\label{eq:(A5)}
\end{align}
Therefore, from Eqs.~\eqref{eq:(A2)}, \eqref{eq:(A3)}, \eqref{eq:(A4)},
and~\eqref{eq:(A5)}, we have the flatness (setting
$n-\Hat{1}-\Hat{2}-\Hat{3}\to n$)
\begin{equation}
   \Delta_1z_{23}(n)
   +\Delta_2z_{31}(n)
   +\Delta_3z_{12}(n)=0\bmod q.
\label{eq:(A6)}
\end{equation}
This shows in general
\begin{equation}
   \frac{1}{2}\sum_{\nu,\rho,\sigma}\varepsilon_{\mu\nu\rho\sigma}
   \Delta_\nu z_{\rho\sigma}(n)=0\bmod q.
\end{equation}

\section{Use of the non-commutative differential calculus in lattice Abelian
gauge theory~\cite{Fujiwara:1999fi}}
\label{sec:B}
In this appendix, we explain that the particular shift in the lattice
coordinate appearing in~Eqs.~\eqref{eq:(3.11)} and~\eqref{eq:(4.4)}
(and also~\eqref{eq:(4.5)}) is naturally understood from the non-commutative
differential calculus~\cite{Dimakis:1992pk} in lattice Abelian gauge
theory on the hypercubic lattice~\cite{Fujiwara:1999fi}.

We define a $k$-form~$f(n)$ on the lattice~$\Lambda$ by
\begin{equation}
   f(n)\equiv
   \frac{1}{k!}\sum_{\mu_1,\dotsc,\mu_k}
   f_{\mu_1\dotsb\mu_k}(n)dx_{\mu_1}\dotsb dx_{\mu_k},
\label{eq:(B1)}
\end{equation}
where $dx_\mu dx_\nu=-dx_\nu dx_\mu$. The exterior derivative on the lattice is
defined by the forward difference,
\begin{equation}
   df(n)\equiv\frac{1}{k!}
   \sum_{\mu,\mu_1,\dotsc,\mu_k}\Delta_\mu
   f_{\mu_1\dotsb\mu_k}(n)dx_\mu dx_{\mu_1}\dotsb dx_{\mu_k}.
\label{eq:(B2)}
\end{equation}
This is nilpotent, $d^2=0$.

The essence of the non-commutative differential calculus is the rule,
\begin{equation}
   dx_\mu f_{\mu_1\dotsb\mu_k}(n)=f_{\mu_1\dotsb\mu_k}(n+\Hat{\mu})dx_\mu.
\label{eq:(B3)}
\end{equation}
That is, the differential form and a function on the lattice do not simply
commute and the exchange accompanies a shift of the coordinate. If one accepts
this formal rule, one finds that the Leibniz rule of the exterior derivative,
\begin{equation}
   d[f(n)g(n)]=df(n)\cdot g(n)+(-1)^kf(n)dg(n),
\label{eq:(B4)}
\end{equation}
holds even with the lattice difference~\eqref{eq:(B2)}.

With the above understanding, for a two-form
\begin{equation}
   f(n)=\frac{1}{2}\sum_{\mu,\nu}f_{\mu\nu}(n)dx_\mu dx_\nu,
\label{eq:(B5)}
\end{equation}
the wedge product yields
\begin{align}
   f(n)f(n)
   &=\frac{1}{4}
   \sum_{\mu,\nu,\rho,\sigma}
   f_{\mu\nu}(n)f_{\rho\sigma}(n+\Hat{\mu}+\Hat{\nu})
   dx_\mu dx_\nu dx_\rho dx_\sigma
\notag\\
   &=\frac{1}{4}
   \sum_{\mu,\nu,\rho,\sigma}\varepsilon_{\mu\nu\rho\sigma}
   f_{\mu\nu}(n)f_{\rho\sigma}(n+\Hat{\mu}+\Hat{\nu})
   dx_1dx_2dx_3dx_4.
\label{eq:(B6)}
\end{align}
After removing the volume form~$dx_1dx_2dx_3dx_4$ from this, we find the
structure in~Eqs.~\eqref{eq:(3.11)} and~\eqref{eq:(4.4)}. This shows that the
structure appearing in~Eqs.~\eqref{eq:(3.11)} and~\eqref{eq:(4.4)} is rather
natural in lattice Abelian gauge theory.

Now, in terms of the differential forms,
\begin{align}
   z^{(2)}(n)&\equiv\frac{1}{2}\sum_{\mu,\nu}z_{\mu\nu}(n)dx_\mu dx_\nu,\qquad
   z^{(1)}(n)\equiv\sum_\mu z_\mu(n)dx_\mu,
\notag\\
   N(n)&\equiv\frac{1}{2}\sum_{\mu,\nu}N_{\mu\nu}(n)dx_\mu dx_\nu,
\label{eq:(B7)}
\end{align}
the $\mathbb{Z}_q$ one-form gauge transformation~\eqref{eq:(2.21)} is written
as
\begin{equation}
   z^{(2)}(n)\to z^{(2)}(n)+dz^{(1)}(n)+qN(n)
\label{eq:(B8)}
\end{equation}
and, using the Leibniz rule~\eqref{eq:(B4)} and the nilpotency~$d^2=0$,
\begin{align}
   &\sum_{n\in\Lambda}z^{(2)}z^{(2)}
\notag\\
   &\to
   \sum_{n\in\Lambda}z^{(2)}z^{(2)}
   +\sum_{n\in\Lambda}\left(
   -dz^{(2)}z^{(1)}+z^{(1)}dz^{(2)}
   \right)
\notag\\
   &\qquad{}
   +q\sum_{n\in\Lambda}
   \left(
   z^{(2)}N+Nz^{(2)}+qNN+dz^{(1)}N+Ndz^{(1)}\right)
\notag\\
   &\qquad{}
   +\sum_{n\in\Lambda}
   d\left(z^{(2)}z^{(1)}+z^{(1)}z^{(2)}+z^{(1)}dz^{(1)}\right).
\label{eq:(B9)}
\end{align}
In this expression, we can discard the last ``surface term'' because the fields
$z_{\mu\nu}(n)$, $z_\mu(n)$, and~$N_{\mu\nu}(n)$ are \emph{single-valued} on the
lattice.\footnote{Note that the conditions in~Eq.~\eqref{eq:(2.20)}
and~$0\leq z_\mu(n)<q$ uniquely determine these fields on the lattice.} In
terms of the components, we thus have
\begin{align}
   &\sum_{n\in\Lambda}
   \sum_{\mu,\nu,\rho,\sigma}\varepsilon_{\mu\nu\rho\sigma}
   z_{\mu\nu}(n)z_{\rho\sigma}(n+\Hat{\mu}+\Hat{\nu})
\notag\\
   &\to
   \sum_{n\in\Lambda}
   \sum_{\mu,\nu,\rho,\sigma}\varepsilon_{\mu\nu\rho\sigma}
   z_{\mu\nu}(n)z_{\rho\sigma}(n+\Hat{\mu}+\Hat{\nu})
\notag\\
   &\qquad{}
   +\sum_{n\in\Lambda}
   \sum_{\mu,\nu,\rho,\sigma}\varepsilon_{\mu\nu\rho\sigma}
   \left[
   -2\Delta_\mu z_{\nu\rho}(n)z_\sigma(n+\Hat{\mu}+\Hat{\nu}+\Hat{\rho})
   +2z_\mu(n)\Delta_\nu z_{\rho\sigma}(n+\Hat{\mu})
   \right]
\notag\\
   &\qquad{}
   +q\sum_{n\in\Lambda}
   \sum_{\mu,\nu,\rho,\sigma}\varepsilon_{\mu\nu\rho\sigma}
   \{z_{\mu\nu}(n)N_{\rho\sigma}(n+\Hat{\mu}+\Hat{\nu})
   +N_{\mu\nu}(n)z_{\rho\sigma}(n+\Hat{\mu}+\Hat{\nu})
\notag\\
   &\qquad\qquad\qquad\qquad\qquad\qquad{}
   +qN_{\mu\nu}(n)N_{\rho\sigma}(n+\Hat{\mu}+\Hat{\nu})
\notag\\
   &\qquad\qquad\qquad\qquad\qquad\qquad{}
   +2\Delta_\mu z_\nu(n)N_{\rho\sigma}(n+\Hat{\mu}+\Hat{\nu})
   +2N_{\mu\nu}(n)\Delta_\rho z_\sigma(n+\Hat{\mu}+\Hat{\nu})
   \}
\label{eq:(B10)}
\end{align}
under the $\mathbb{Z}_q$ one-form gauge transformation. Since
$(1/2)\sum_{\nu,\rho,\sigma}\varepsilon_{\mu\nu\rho\sigma}
\Delta_\nu z_{\rho\sigma}(n)=0\bmod q$ (the flatness), from this expression,
it is obvious that the shift of~$\sum_{n\in\Lambda}
\sum_{\mu,\nu,\rho,\sigma}\varepsilon_{\mu\nu\rho\sigma}
z_{\mu\nu}(n)z_{\rho\sigma}(n+\Hat{\mu}+\Hat{\nu})$ under the gauge transformation
is~$4q\mathbb{Z}$.\footnote{We confirmed that the shift actually can take a
value in~$4q\mathbb{Z}$ by a numerical experiment. When $dz^{(2)}=dN=0$
(strictly zero \emph{not\/} modulo~$q$), we can show that the shift
is~$8q\mathbb{Z}$ instead of~$4q\mathbb{Z}$ by employing the argument
in~Ref.~\cite{Fujiwara:2000wn}.}



%



\let\doi\relax









\end{document}